\documentstyle[prb,aps,preprint]{revtex}

\begin{document}
\draft
\title{Coupling to spin fluctuations from conductivity scattering rates}
\author{E. Schachinger$^1$ and J.P. Carbotte$^2$}
\address{$^1$Institut f\"ur Theoretische Physik, Technische Universit\"at
Graz, A-8010 Graz, Austria\\
$^2$Department of Physics and Astronomy, McMaster University,
Hamilton, Ont. L8S 4M1, Canada}
\date{\today}
\maketitle
\begin{abstract}
A recent analysis of optical conductivity data which has provided
evidence for coupling of the charge carriers to the 
$41\,$meV spin resonance seen in the superconducting state of
optimally doped YBa$_2$Cu$_3$O$_{6.95}$ (Y123), is extended to
other systems. We find that the corresponding
spin resonance is considerably broader in Tl$_2$Sr$_2$CuO$_{8+\delta}$
(Tl2201) and YBa$_2$Cu$_4$O$_8$ (Y124) than it is in
Bi$_2$Sr$_2$CaCu$_2$O$_{8+\delta}$ (Bi2212) and there is no
resonance in overdoped Tl2201 with $T_c = 23\,$K.
The effective charge-spin spectral density is
temperature dependent and contains
feedback effects that further stabilize superconductivity as $T$
is reduced.
\end{abstract}
\newpage
For a conventional electron-phonon system an isotropic, on the Fermi
surface, spectral density can be introduced which is essentially
temperature independent below $T_c$.\cite{Carb1} This spectral
density, $\alpha^2F(\omega)$, can be determined from tunneling
data in the superconducting state and has been used with great
success to understand the deviations from BCS universal laws
observed in many conventional superconductors \cite{Carb1}.
In principle, information on $\alpha^2F(\omega)$ can also be
obtained through inversion of optical data although, to our
knowledge, this has only been accomplished for Pb.\cite{FarmTim}

Recently Marsiglio {\it et al.}\cite{Mars1} introduced a 
dimensionless function $W(\omega)$ which
is defined as the second derivative of the normal state optical
scattering rate 
$\tau^{-1}(\omega) = (\Omega^2_p/4\pi)\Re{\rm e}\sigma_N^{-1}(\omega)$
 multiplied by frequency $\omega$. Here $\Omega_p$ is the plasma
frequency and $\sigma_N(\omega)$ the normal state optical
conductivity.  Specifically,
\begin{equation}
  W(\omega) = {1\over 2\pi}{d^2\over d\omega^2}\left[
  \omega\over\tau(\omega)\right]
  \label{eq:1}
\end{equation}
which follows directly from experiment, provided the data on
$\sigma_N(\omega)$ is sufficiently accurate that a meaningful
second derivative can be taken, possibly after smoothing.
Marsiglio {\it et al.}\cite{Mars1} made the very important observation
that within the phonon range
$W(\omega)\simeq \alpha^2F(\omega)$ at least for those
spectral densities studied. Beyond the phonon
range $W(\omega)$ can be negative but this does not distract from the
fact that $W(\omega)$ can be used to get the shape and magnitude of
$\alpha^2F(\omega)$.
Application of Eq.~(\ref{eq:1}) to the normal state conductivity
of K$_3$C$_{60}$ gave an
$\alpha^2F(\omega)$ (provided negative regions in $W(\omega)$
are simply ignored)
in excellent agreement with incoherent inelastic neutron scattering
data\cite{Neut,Corr} on the phonon frequency distribution
$F(\omega)$ and gave sufficient coupling strength to obtain the measured value
of $T_c$. This leaves little doubt that K$_3$C$_{60}$ is an
$s$-wave, electron-phonon superconductor, even though correlation
effects are likely to be quite important.\cite{Corr}

More recently Carbotte {\it et al.}\cite{Carb2} have extended the method
of Ref.~3 to spin fluctuation exchange systems and to the 
superconducting state with $d$-wave symmetry. The charge
carriers are coupled to the spin fluctuations through the
spin susceptibility which is
strongly peaked at $(\pi,\pi)$ in the two dimensional 
Brillouin zone of the CuO$_2$ plane
of the high $T_c$ oxides.\cite{Mont2}
In this case the momentum dependence of the interaction is very
important and cannot be pinned to the Fermi surface\cite{Branch}
and there are cold and hot spots. Nevertheless, the resulting
in-plane infrared conductivity is isotropic for tetragonal systems
and Eq.~(\ref{eq:1}) can still be applied and the resulting
$W(\omega)$ interpreted as
an effective spectral density for the electron-spin fluctuation
exchange interaction.\cite{Branch2} In contrast to the
electron-phonon case this effective interaction resides in the
system of electrons and, due to correlation effects,
can be temperature dependent. In particular, it can undergo major
changes when the electrons
condense into the superconducting state. Such feedback effects
are generic to any electronic mechanism.\cite{Bonn,Nuss,Schach}
They have been studied theoretically within a Hubbard model by
Dahm and Tewordt\cite{Dahm} who also review the work of others.

Optical conductivity calculations for a $d$-wave
superconductor\cite{Schach,Munz} within a spin fluctuation mechanism
by Carbotte {\it et al.}\cite{Carb2}
established, that $W(\omega)$ of Eq.~(\ref{eq:1}) still
gives a good approximation to
the spectral density $I^2\chi(\omega)$ provided it is divided by two
and shifted by the gap $\Delta_0$. Calculations of $W(\omega)$ 
from the data of Basov {\it et al.}\cite{Basov}
in optimally doped Y123, revealed strong coupling of the
charge carriers to the $41\,$meV
spin resonance seen below $T_c$ in spin polarized
inelastic neutron scattering experiments.\cite{Ross,Bourg} The
coupling to this resonance
was found to be large enough to stabilize the
observed superconducting state. For underdoped Y123 the spin resonance
remains in the optics even above $T_c$ up to a pseudogap temperature,
in agreement with the neutron work by Dai {\it et al.}\cite{Dai}
A quantitative analysis is not attempted in this case however, because
of the added complications of the pseudogap.

Here we extend our previous work\cite{Carb2} to other materials and,
in contrast to what was done in Ref.~6 we proceed here without any
reference to neutron data.
In Fig.~\ref{f1} we show results for the coupling to the spin
resonance in Y124 ($T_c = 82\,$K, solid line), Tl2201 ($T_c = 90\,$K, dashed
line), and Bi2212 ($T_c = 90\,$K, dotted line) derived from optical
data measured at $T=10\,$K.
These results were obtained from a direct application of Eq.~(\ref{eq:1})
to the optical data
of Puchkov {\it et al.}\cite{Puch} Shifting by the gap
which is determined by the method discussed in detail later on,
the resonances
are at 38, 43\cite{Schach1}, and $46\,$meV respectively with a considerably larger
width in the first two than in Bi2212. On the other hand,
the spin resonance in Bi2212 was observed by Fong {\it et al.}\cite{Fong}
at $43\,$meV using neutron scattering.
No neutron data exist, to our knowledge, for Y124 and Tl2201 and, therefore,
our results represent a falsifiable prediction. Another
prediction that we develop later is that, overdoped Tl2201 ($T_c = 23\,$K)
will show no resonance.

To extract more information from optical data we need to
consider a more specific mechanism, namely spin
fluctuation exchange.\cite{Mont2,Millis1} At the simplest level in
the normal state, we describe the corresponding spectral density by a two 
parameter form
\begin{equation}
  I^2\chi(\omega) = I^2{\omega\omega_{SF}\over\omega^2+\omega_{SF}^2},
  \label{eq:2}
\end{equation}
where $I^2$ is the coupling between spin excitations and the
charge carriers and $\omega_{SF}$ sets the 
energy scale for the spin fluctuations.
Both parameters can be derived from a fit
to the normal state optical scattering rates as a function of frequency.
A fit for Tl2201 ($T_c = 90\,$K) is shown
in the top frame of Fig.~\ref{f2}. The fit to the $T=300\,$K data with
$\omega_{SF} = 100\,$meV and a high energy cutoff at $400\,$meV
is excellent and lowering $\omega_{SF}$ to
$30\,$meV does not give an acceptable fit. The formalism we
use to relate spectral density to conductivity is standard\cite{Mars1}
and $\sigma_N(\omega)$ follows from a knowledge of the
self energy $\Sigma(\omega)$. 
As a check on the accuracy of the inversion procedure
embodied in Eq.~(\ref{eq:1}), we show in the central frame of Fig.~\ref{f2}
our results for the function
$W(\omega)$ obtained from our theoretical normal state
optical scattering rate $\tau^{-1}(\omega)$ based on our input spectral 
density $I^2\chi(\omega)$ given in
Eq.~(\ref{eq:2}) and shown as the grayed squares.
We see, that at $T = 10\,$K the inversion matches
almost perfectly the input spectral density except
for small wiggles in the inverted curve
(solid line). This excellent agreement between $W(\omega)$ and
$I^2\chi(\omega)$ is not limited to simple, smooth forms. In the
bottom frame of Fig.~\ref{f2} we show results obtained for a
structured spectrum, namely a spectrum which is proportional to the
one used by Schachinger and Carbotte\cite{Schach1} to analyze the
optical properties of superconducting Tl2201. Except for some
oscillations at higher frequencies the normal state $W(\omega)$ (solid curve)
is close to the input spectral function $I^2\chi(\omega)$ (grayed squares).

Normal state conductivity data is not available at
low temperatures in the high $T_c$ oxides and it is necessary to
devise an inversion technique which applies in the superconducting
state. Also, the spectral density can depend on temperature and on
the state of the system. This requires a formalism which relates the
spectral density $I^2\chi(\omega)$ to the superconducting
state conductivity. This was provided in the work of Schachinger
{\it et al.}\cite{Schach} who calculated the conductivity of a $d$-wave
superconductor within an Eliashberg formalism. As previously stated,
using this formalism  Carbotte {\it et al.}\cite{Carb2} established
that in this case $W(\omega)/2$ agrees fairly well with $I^2\chi(\omega)$
provided it is shifted by the gap amplitude $\Delta_0$.
This is shown clearly in Fig.~\ref{f3} which
is similar to the bottom frame of Fig.~\ref{f2} except that now the
superconducting state conductivity has been employed and the material
is Bi2212 with $T_c = 90\,$K rather than Tl2201. The grayed
squares are the input spectral density shifted by the 
theoretical gap $\Delta_0 = 28\,$meV
and the dashed line are the results for the inversion
$W(\omega)/2$ vs.\ $\omega$ based on
the calculated $\sigma_S(\omega)$. A simple $d$-wave gap model
was used, and a parameter $g$ introduced giving
the relative weight of the spin
fluctuation spectral density in the gap channel as compared to its
value in the renormalization channel.
Details can be found in Ref.~12.
For Bi2212, $g=0.725$, gives the measured value of $T_c$ when the normal
state spectral density of Eq.~(\ref{eq:2}) is used in the linearized
self energy equations at $T=T_c = 90\,$K. This value of $g$, which is
considerably less than one, could be interpreted as an indication
that a second, subdominant scattering mechanism (for example
phonons) is also operative. The theoretical gap, on the other hand, is
calculated from the solution of the $d$-wave Eliashberg equations\cite{Schach}
for a temperature $T = 10\,$K and is defined as the peak in the
quasiparticle density of states. It is to be noted that this gap is
a bit smaller than the gap of $31\,$meV suggested from the inversion
data of Fig.~\ref{f1} for Bi2212 (dotted line) using the experimentally
observed position of the resonance peak at $43\,$meV.\cite{Fong}
Nevertheless, the agreement is excellent and the theoretical value of
$28\,$meV is within the experimentally observed range.\cite{Hasegawa}

The agreement between $W(\omega)/2$ and $I^2\chi(\omega)$ in the top
frame of Fig.~\ref{f3}
as to size and shape of the main peak is
excellent. However, a negative piece is introduced in $W(\omega)$ right 
above the spin resonance
peak which is not part of the spectral density. Nevertheless,
at higher energies, $W(\omega)/2$ does recover and shows long tails extending 
to several
$100\,$meV although they are underestimated. Additional evidence
for the existence of this high energy background is found from
our fit to the normal state data shown in the bottom frame of
Fig.~\ref{f3}. The grayed lines give the optical scattering rate in
Bi2212 at $T=300\,$K. The dashed curve is the fit to this data (solid line)
and gives a normal state
spin fluctuation frequency $\omega_{SF} = 100\,$meV in
Eq.~(\ref{eq:2}) and an area under $I^2\chi(\omega)$ of $95\,$meV.
From application of
Eq.~(\ref{eq:1}) to the superconducting state data we have
already established the existence of coupling of charge carriers
to a resonance peak as seen in Fig.~\ref{f1} which gives its size and
position in energy and this is reproduced as the solid curve in the
top frame of Fig.~\ref{f3}. To get the
superconducting state spectral density (grayed squares of the top
frame in Fig.~\ref{f3}) the low frequency part of the normal state
response is replaced by the resonant peak.

There is no known sum rule
on the spectral weight $I^2\chi(\omega)$ and we find that the area
under this function increases from $95\,$meV in the normal state
to $115\,$meV in the superconducting state. The increase is due to
the appearance of the spin resonance. Part
of this spectral weight could come from a transfer from higher
energies but our resolution at such energies is not sufficient
to confirm this.
In the bottom frame of Fig.~\ref{f3} we show
the fit to the superconducting state optical scattering rate
obtained from our model $I^2\chi(\omega)$.
The agreement is very good and since no new
parameters were introduced
to obtain the black dashed curve which agrees remarkably well with
the black solid curve in the region $0\le\omega\le 250\,$meV,
this is taken to be a strong consistency check on our
work.

We extend our analysis to the material Y124 $(T_c = 81\,{\rm K})$
where we predict from Fig.~\ref{f1} a spin resonance to exist at
$38\,$meV. Moreover, this spin resonance is much broader than the
one observed in Bi2212 of Y123. Results are presented in Fig.~\ref{f4}.
The top frame of this figure demonstrates the agreement between
$W(\omega)/2$ and $I^2\chi(\omega)$ which was shifted by the
theoretical gap $\Delta_0 = 24\,$meV which is another prediction
of our calculations as, to our knowledge, no experimental data
exist for this material. The bottom frame of Fig.~\ref{f4}
presents our comparison between experimental and theoretical
optical scattering rates. As in the case of Bi2212 the normal
state scattering rate (grayed lines) at $T=300\,$K gives evidence
for the existence of a high energy background as the experimental
data (solid line) are best fit by a spin fluctuation spectrum
of the type described by Eq.~(\ref{eq:2}) with $\omega_{SF} = 80\,$meV
and a high energy cutoff of $400\,$meV (dashed line). The black lines
compare the theoretical results (dashed line) to experiment (solid
line) in the superconducting state at $T=10\,$K. The signature
of the spin resonance, the sharp rise in $\tau^{-1}(\omega)$
starting around $50\,$meV is correctly reproduced by theory. For
$\omega > 120\,$meV the experimental scattering rate shows only a
weak energy dependence and the theoretical prediction starts
to deviate from experiment. This is in contrast to our results for
Bi2212 (bottom frame of Fig.~\ref{f3}) and Tl2201\cite{Schach1}
and could be related to the fact that the Y124 sample used by
Puchkov {\it et al.}\cite{Puch} was slightly underdoped, a
situation not covered by our theory.

To conclude, we obtained theoretical gap amplitudes
$\Delta_0 = 24, 26,$ and $28\,$meV for Y124, Tl2201\cite{Schach1}, and Bi2212
respectively. Experimental values are in the range of $30\,$meV for
Bi2212 and $28\,$meV for Tl2201.\cite{Hasegawa} The theoretical
values correspond to ratios $2\Delta_0/k_BT_c$ of
6.8, 6.7, and 7.2, much larger than the BCS value of $\sim 4.3$, and
proves that feedback effects, not present in BCS, stabilize
the superconducting state as $T$ is reduced (a result also
supported by the theoretical study of Dahm and Tewordt\cite{Dahm}).

Next we consider the case of overdoped Tl2201 with
$T_c = 23\,$K. The data of Puchkov {\it et al.}\cite{Puch}
are reproduced as the solid
curves of Fig.~\ref{f5}. The grayed curves are at $T=300\,$K in the
normal state and the black curves apply to the superconducting
state at $T=10\,$K. Nowhere is there a large rapid rise
in the $T=10\,K$ curve at an energy which would
correspond to the sum of $\Delta_0$ plus some resonant
frequency. This is in striking contrast
to the sharp rise seen in Bi2212 and Y124 (bottom frame of Figs.~\ref{f3}
and \ref{f4} solid curves). For this overdoped sample no spin
resonance forms. In fact, a fit of
Eq.~(\ref{eq:2}) with $\omega_{SF}=300\,$meV to the
$T=300\,$K data which gives the grayed dashed curve in good agreement
with the data (grayed, solid line) also
gives the black dashed curve when used in a superconducting
state calculation. The agreement with the solid black curve is quite good. 
No adjustment of any kind was made.
Finally, we note in passing that $\tau^{-1}(\omega)$ stays
finite (but very small) in the limit $\omega\to 0$ in all calculations
presented here.

Optical conductivity data in
Bi2212, Y124, and Tl2201\cite{Schach1} indicate that in the superconducting
state the charge carriers are strongly coupled to a spin resonance 
which forms only in this state. No such resonance is seen in overdoped
Tl2201 with $T_c = 23\,$K. These results confirm that the coupling to the
spin resonance first seen in optimally doped Y123 is a general feature
of several, but not all the high $T_c$ oxides. The feature that 
corresponds to the resonance is an sharp rise in the optical
scattering rate at a frequency equal to the sum of the gap
plus the spin resonance frequency. Inversion of the optical data gives
information on the absolute strength of the coupling between
charge carriers and the spin resonance, and on its width. The
resonance is found to be considerably broader in Tl2201
and Y124 than it is in Bi2212. In the
systems considered here, at $T_c$, there is only coupling to the 
background spin
fluctuations which extend to high energies. Below $T_c$ a
spin resonance forms\cite{Dai} at low $\omega$ and this leads 
to increased coupling to the spin degrees of freedom which
further stabilizes the superconducting state.
This feedback effect leads to a ratio of twice the
gap to $T_c$ of order 6-8, much larger
than the value predicted in weak coupling BCS for a $d$-wave gap
$(\sim 4.3)$.
Overdoped Tl2201 with $T_c = 23\,$K provides an example for which
no spin resonance forms below $T_c$, and this system has optical
properties close to those expected for a Fermi liquid.

Research supported in part by NSERC (Natural Sciences and
Engineering Research Council of Canada) and by CIAR (Canadian
Institute for Advanced Research). We thank D.N.\ Basov for
continued interest in this work and discussions.

\newpage
\begin{figure}
\caption{
The spin resonance obtained from inversion of (supercond.)
optical conductivity data using Eq.~(\ref{eq:1}) for
$W(\omega)$: solid line Y124 ($T_c = 82\,$K),
dashed Tl2201 ($T_c = 90\,$K), and dotted
Bi2212 ($T_c = 90\,$K). Note that the vertical scale is dimensionless.
The position of the resonance peak is shifted by the gap value
on the horizontal scale.}
\label{f1}
\end{figure}
\begin{figure}
\caption{The top frame gives the optical scattering rate at
$T=300\,$K in the normal state of Tl2201 ($T_c = 90\,$K). The
solid curve is experiments, the dashed one a fit with
$\omega_{SF} = 100\,$meV in Eq.~(\ref{eq:2}), and the dotted line
is for $\omega_{SF} = 30\,$meV. The middle frame gives
the input $I^2\chi(\omega)$ (grayed solid squares) used in 
the conductivity calculations which give
$W(\omega)$ (solid line) at $T=10\,$K. 
The bottom frame
gives another model for $I^2\chi(\omega)$ based on Tl2201
(grayed solid squares)
and the corresponding $W(\omega)$ at $T=10\,$K.
}
\label{f2}
\end{figure}
\begin{figure}
\caption{The top frame gives our model for the spin fluctuation
spectral density (displaced by the theoretical gap $\Delta_0 = 28\,$meV)
for Bi2212 in the superconducting state at $T=10\,$K (grayed solid
squares). The dashed line is $W(\omega)/2$ obtained from the calculated
conductivity and the solid line is the
resonant peak of Fig.~\ref{f1} (dotted line) used in constructing the model
$I^2\chi(\omega)$. The bottom frame shows two sets of optical
scattering rates and theoretical fits to these. The solid lines
are experimental and the
dashed lines are the theoretical results. The grayed
lines are for the normal state at $T=300\,$K and the black ones
are for the superconducting state at $T=10\,$K.}
\label{f3}
\end{figure}
\begin{figure}
\caption{The same as Fig.~\ref{f3} but for the material Y124. The
spin fluctuation spectral density $I^2\chi(\omega)$ was displaced
by the theoretical gap $\Delta_0 = 24\,$meV in the top frame.}
\label{f4}
\end{figure}
\begin{figure}
\caption{
The optical scattering rates in
an overdoped sample
of Tl2201 with a $T_c = 23\,$K. The solid lines represent the
experimental data and the dashed lines fits. The grayed curves
apply in the normal state at $T=300\,$K and the black curves
in the superconducting state at $T=10\,$K.
No spin resonant peak is found in this case in contrast to
the three cases shown in Fig.~\ref{f1}.
}
\label{f5}
\end{figure} 
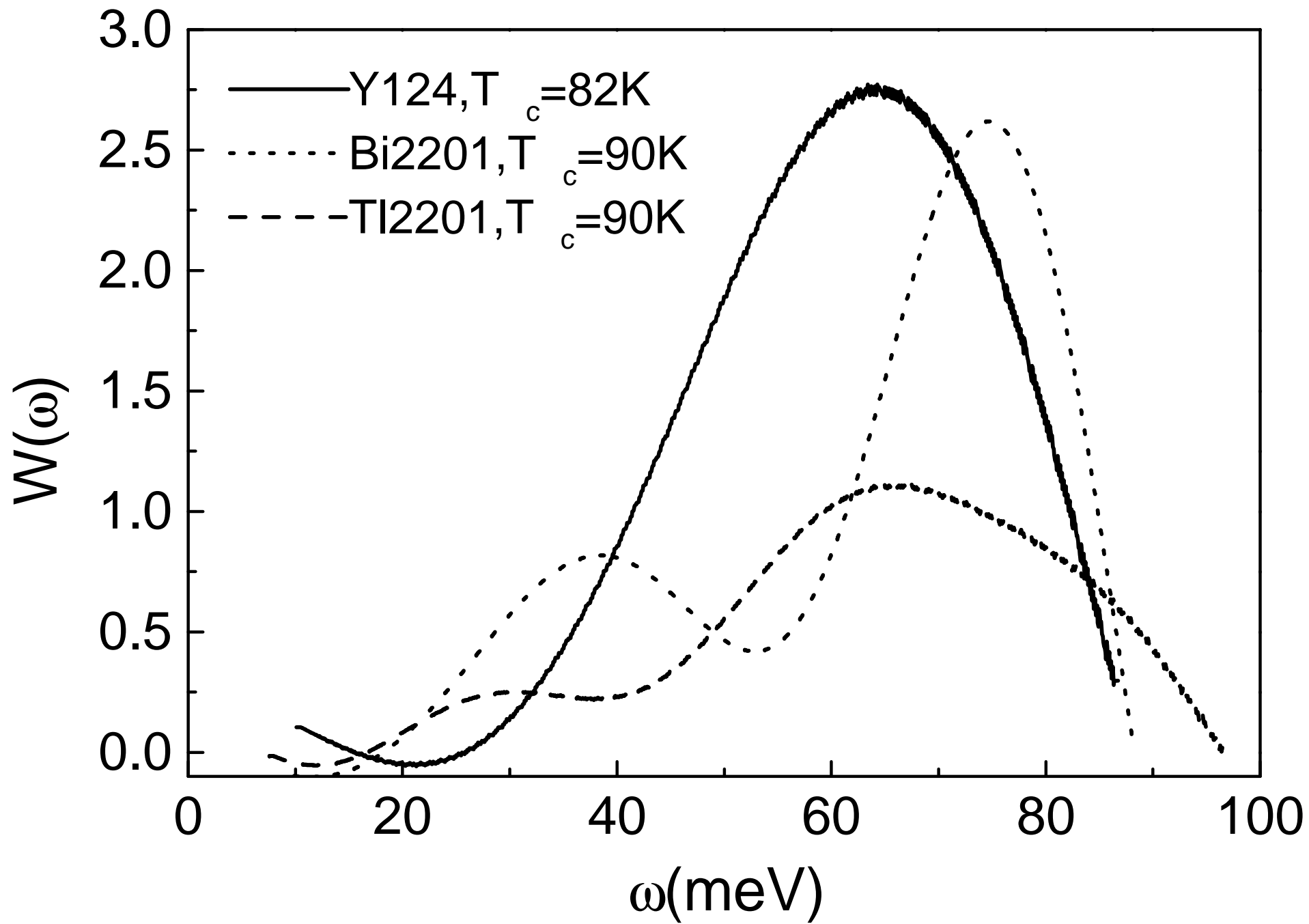
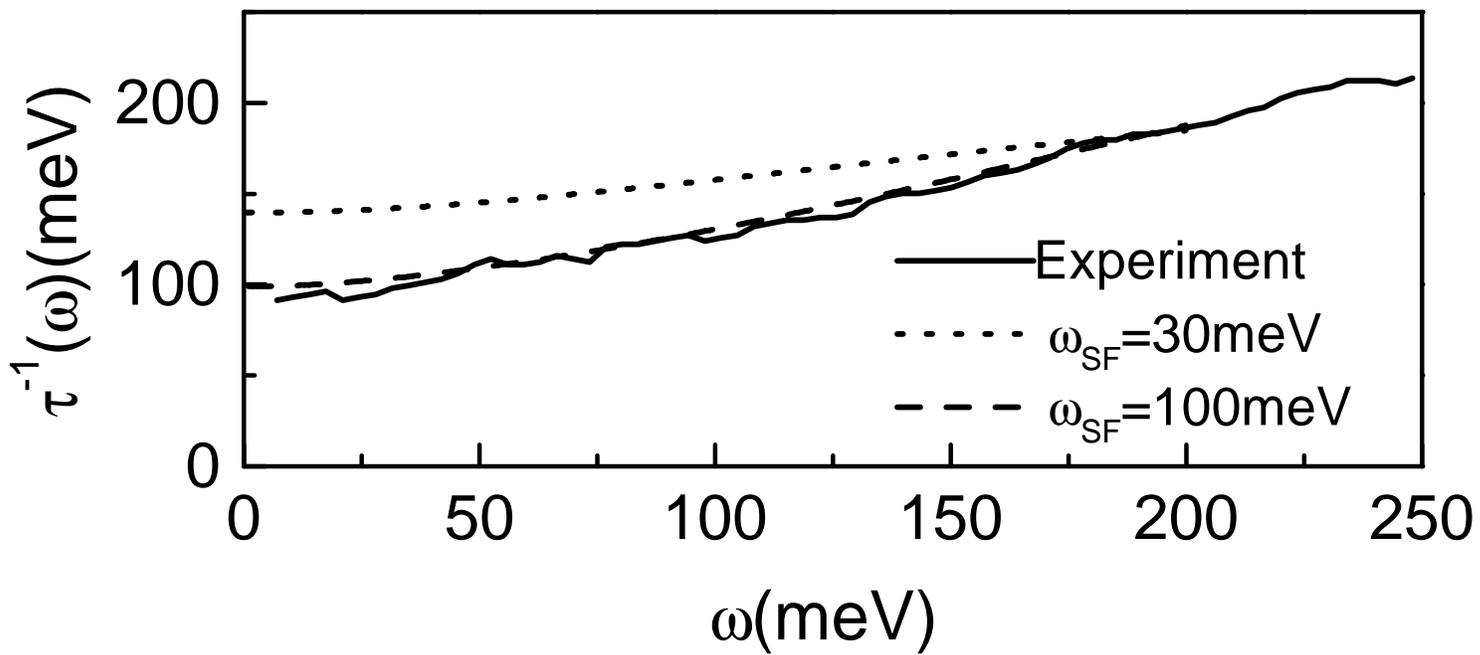
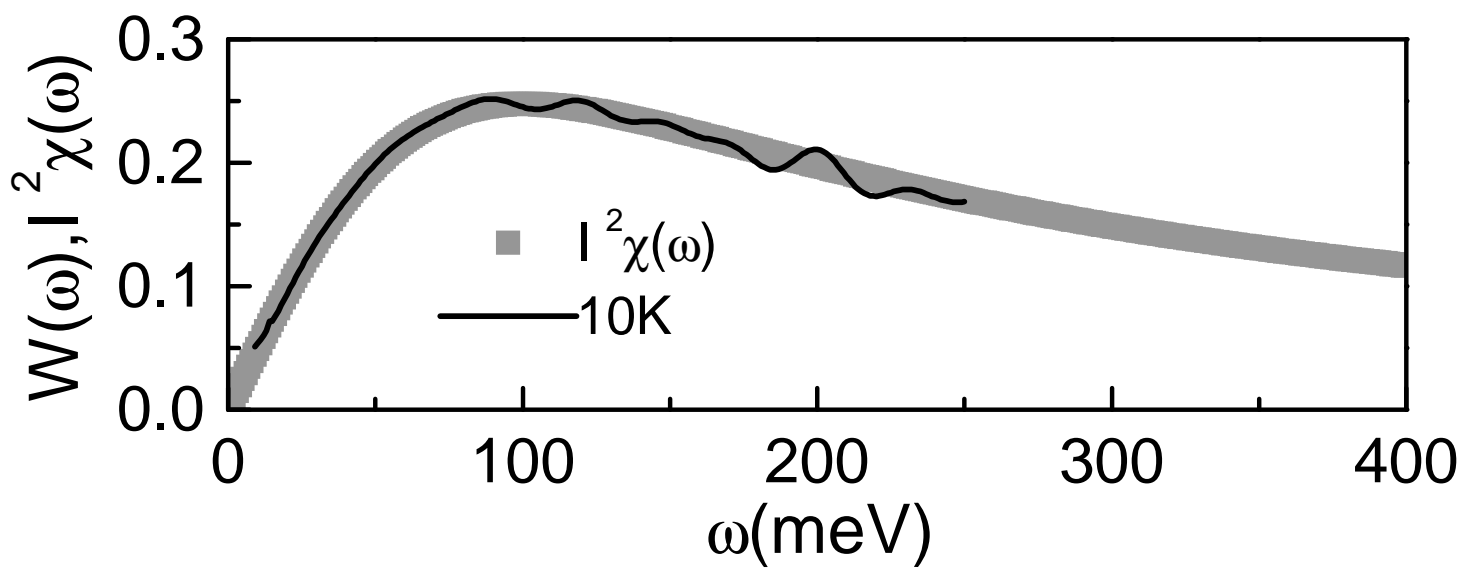
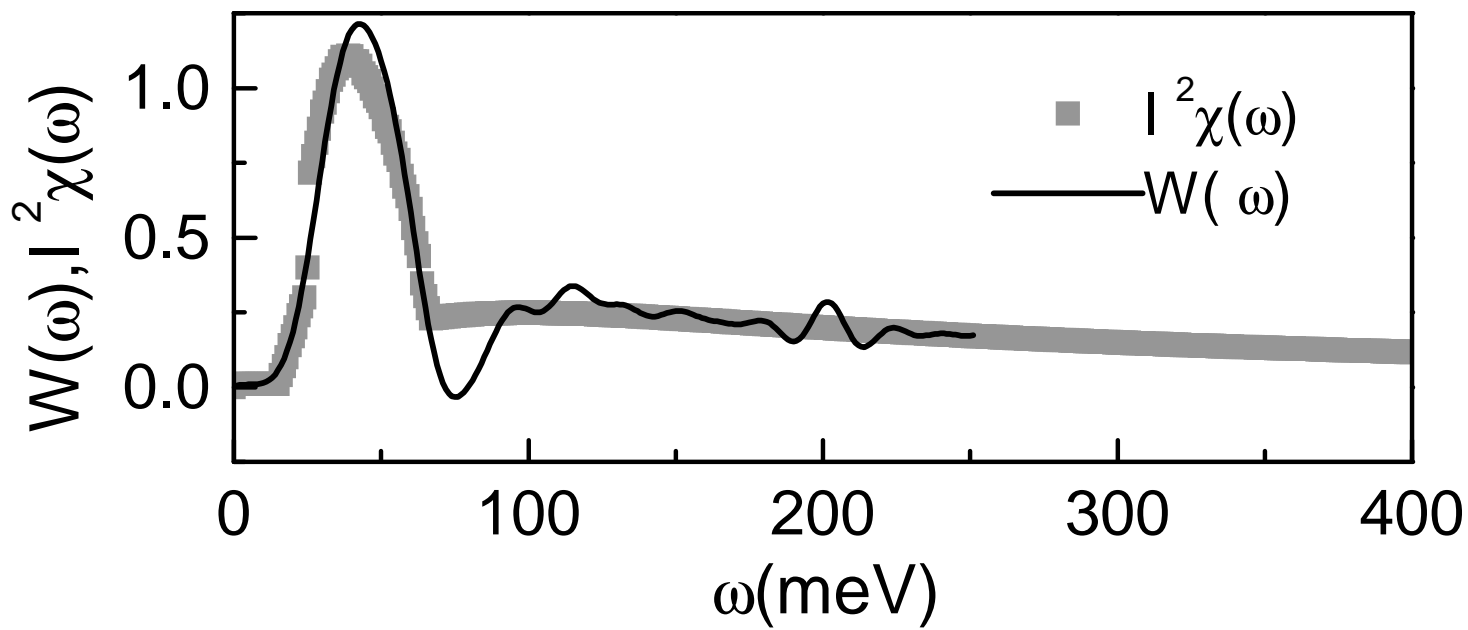
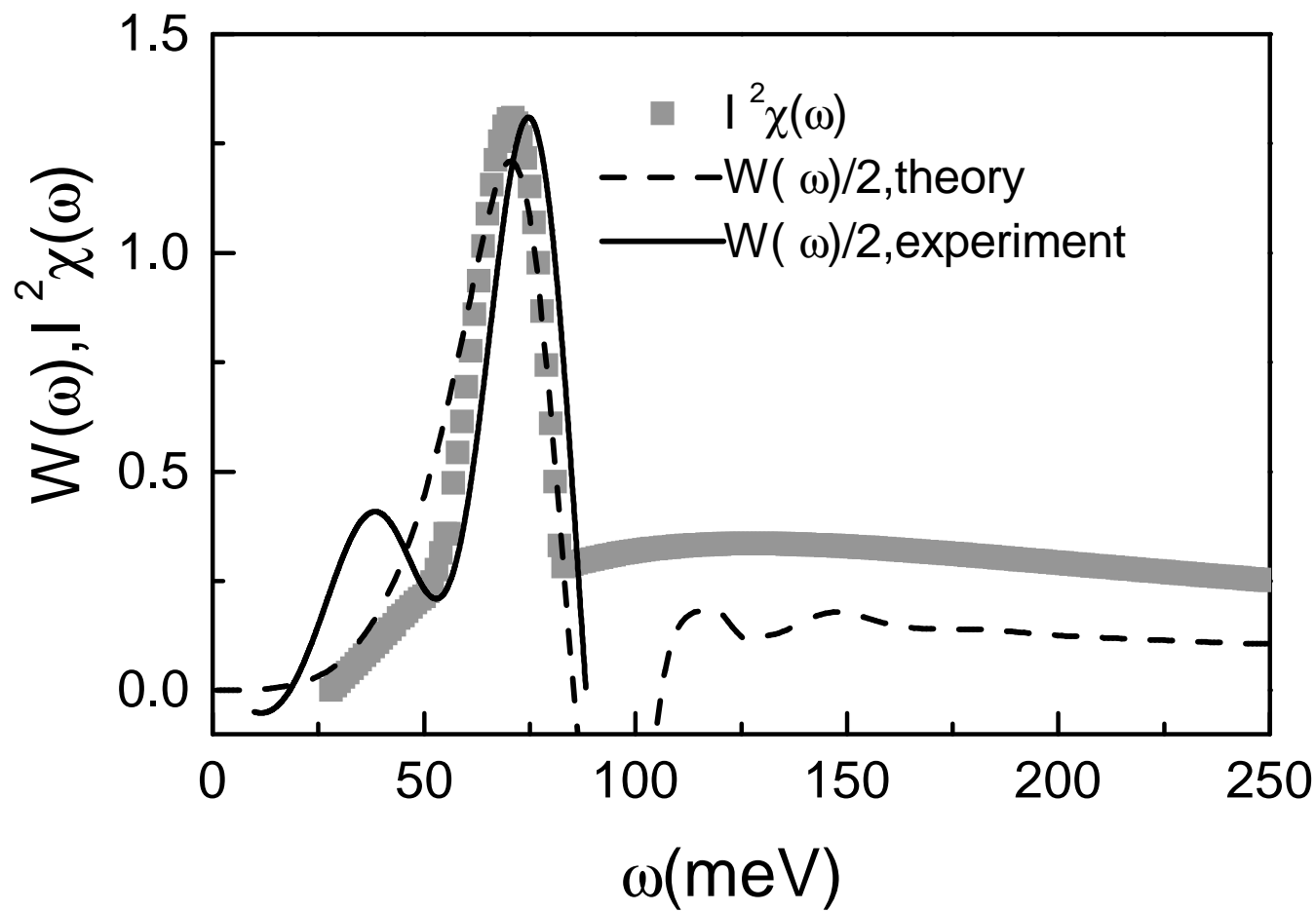
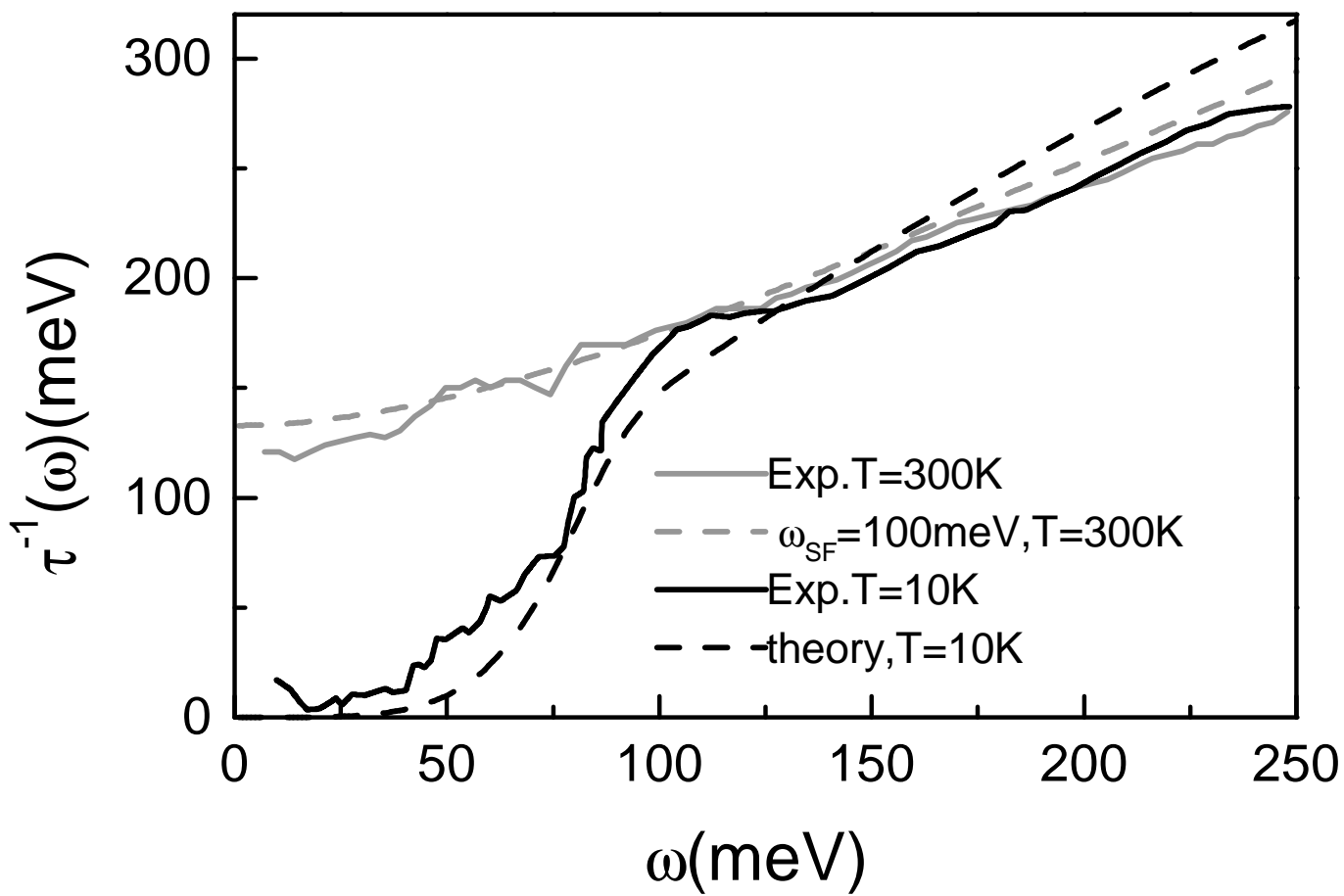
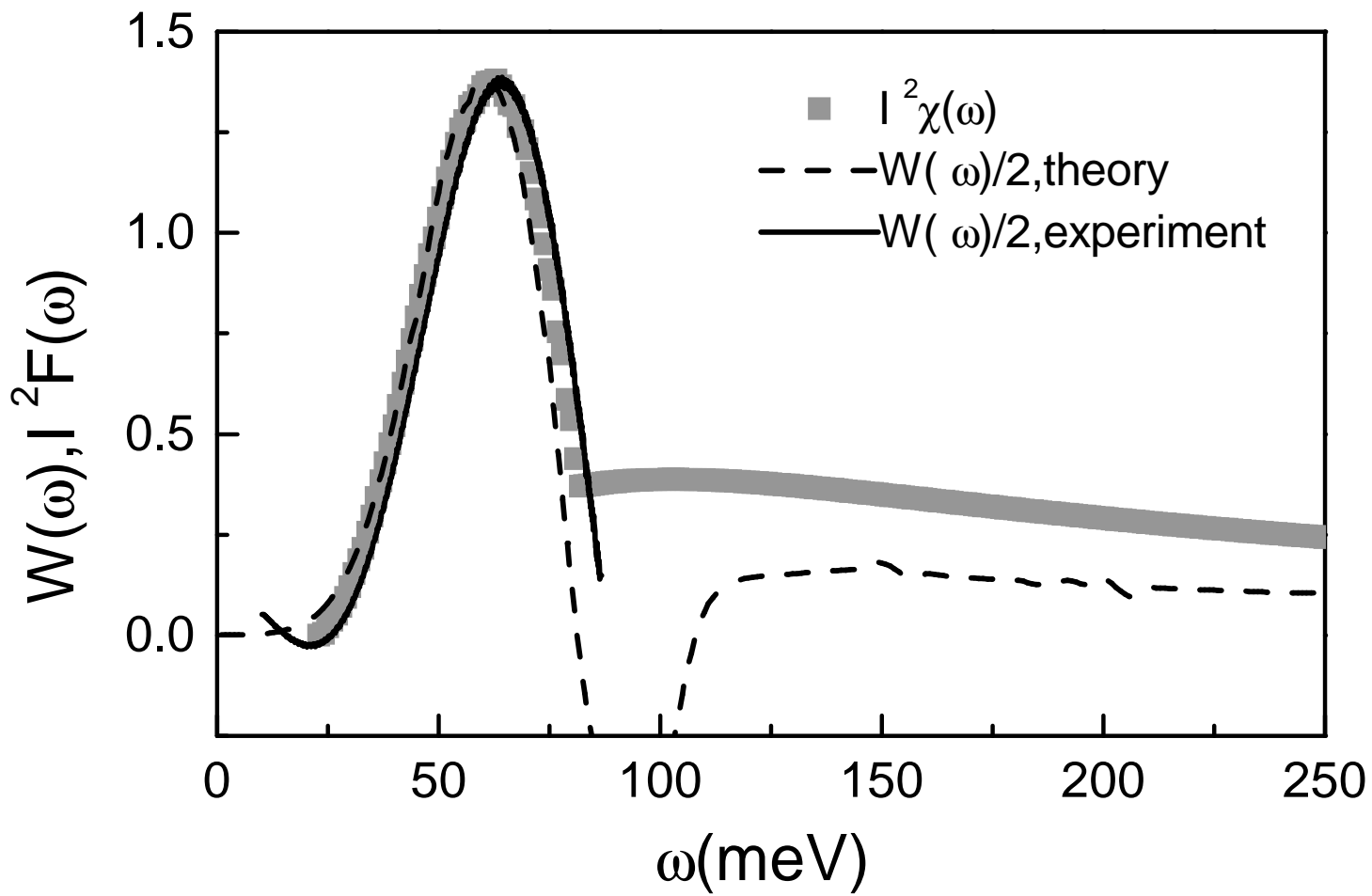
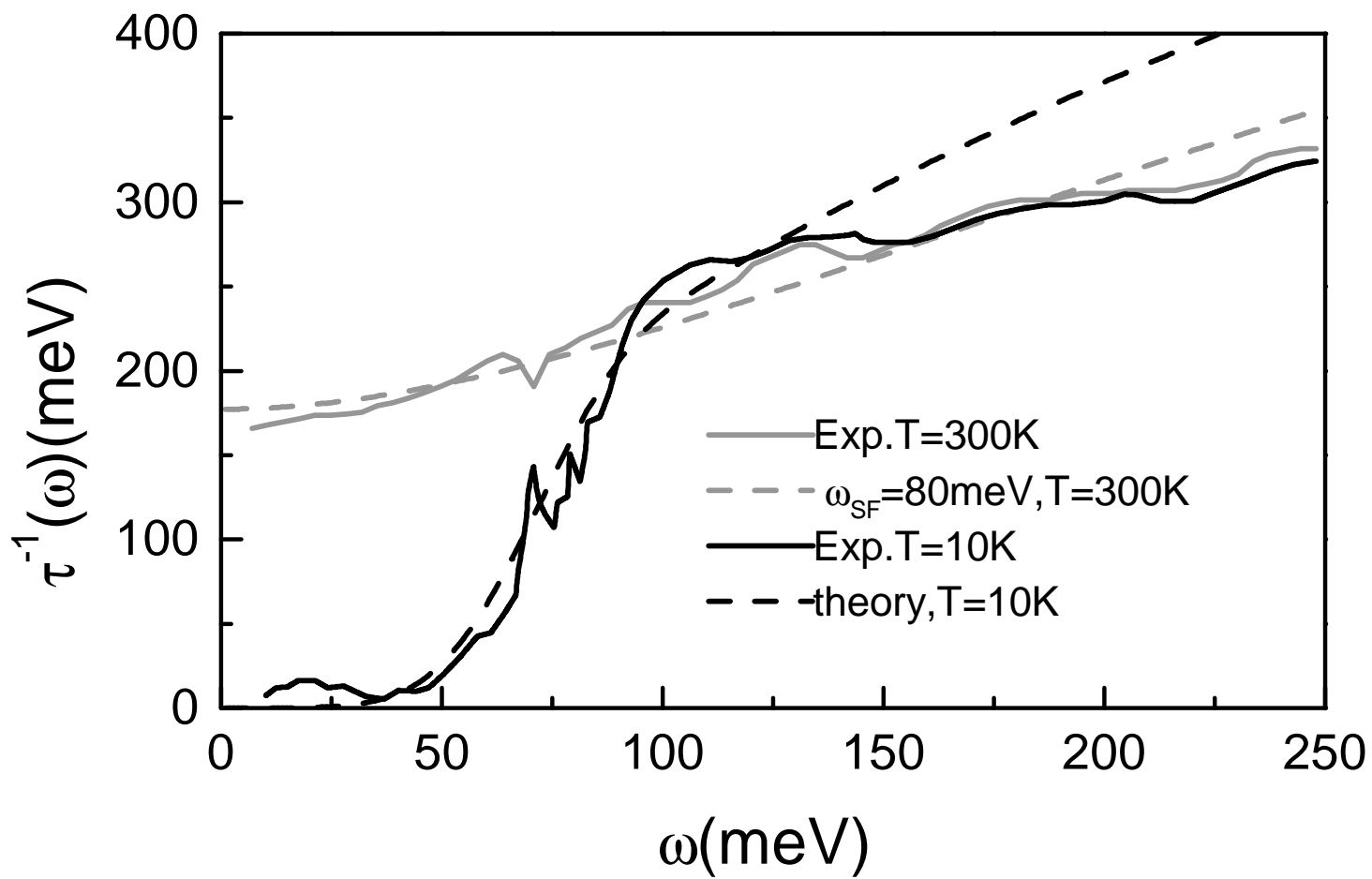
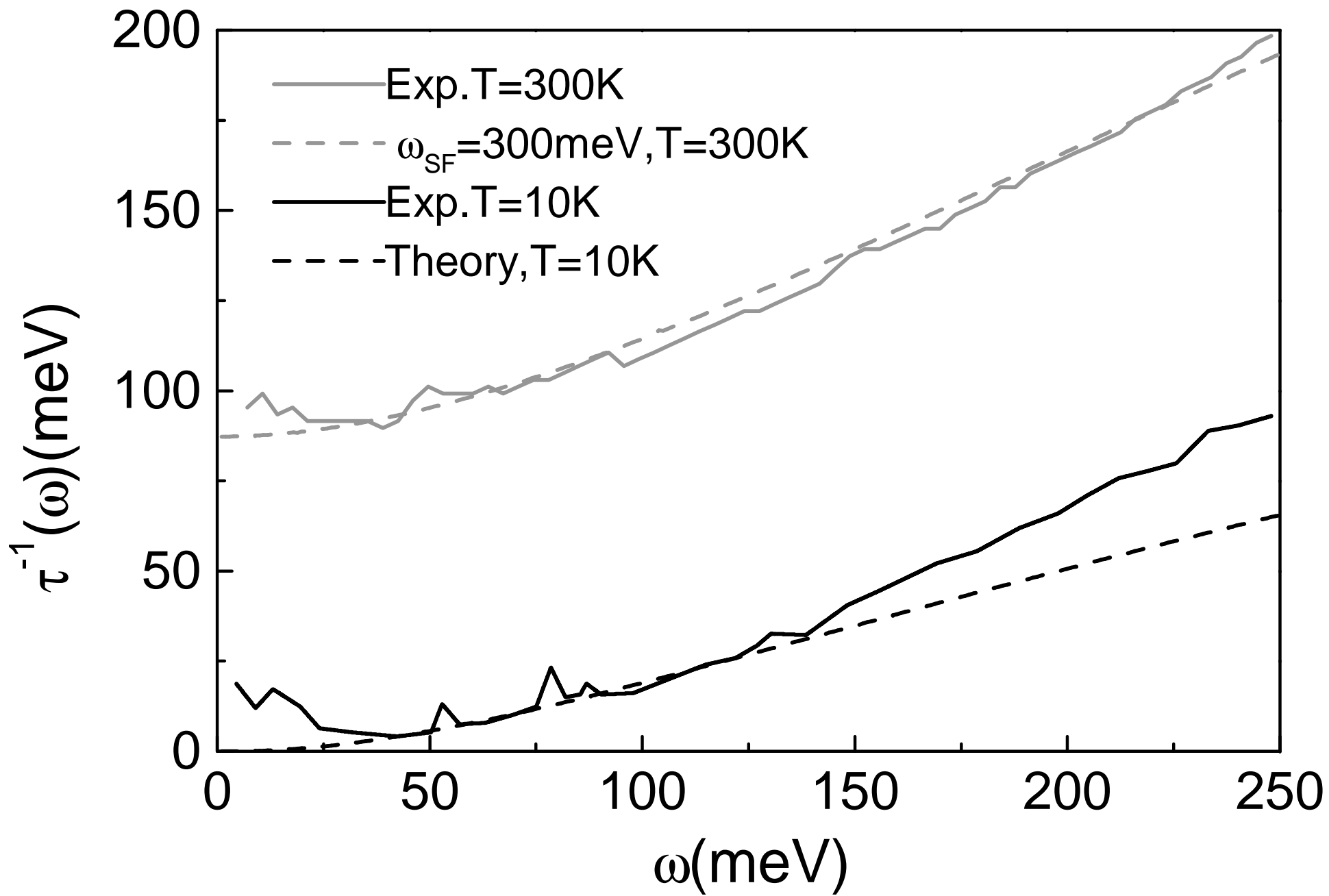
\end{document}